\documentclass[fleqn,twoside]{article}
\usepackage{espcrc2}
\usepackage{epsfig}
\usepackage[figuresright]{rotating}

\newcommand{\AmS}{{\protect\the\textfont2
  A\kern-.1667em\lower.5ex\hbox{M}\kern-.125emS}}
  \title{A solution of a hoary conundrum: the origin and properties of cosmic rays}

\author{A. De R\'ujula\address[MCSD]{Theory Division, CERN,
1211 Geneva 23, Switzerland\\
Physics Department, Boston University, USA}}

\begin{document}
%

%



\begin{abstract}
I discuss a theory of non-solar cosmic rays (CRs) based on a single
type of CR source at all energies. All observed properties
of CRs are predicted in terms of
very simple and completely `standard' physics.  The source of CRs is 
extremely `economical': it has only one parameter to be fitted to 
the enormous ensemble of all of the data. All other inputs are `priors', that is 
theoretical or observational items of information independent of the 
properties of the source of CRs and chosen to lie in their pre-established 
ranges. 

\end{abstract}
\maketitle
%
\section{Introduction}
A couple of years ago, I presented at a large cosmic-ray (CR) conference
a preliminary version of the theory \cite{Florence} I am to summarize here. 
The coffee break subsequent to my talk was attended by hundreds of participants.
A prominent CR theorist, who had heard my talk, told me: {\it `Your theory
is unacceptable. It solves the whole problem'}. He paused to make a large hand 
gesture embracing the surrounding crowd, and added: {\it `What are then
all these people going to do?} Was he just joking? Not that I am convinced
that we have {\it the complete solution} to the problem. But, I shall argue,
we have {\it a} solution that Ockham would have favoured, for the reasons stated in
the abstract.

The {\it standard} theory posits that, up to the knee, CRs are
produced by supernovae (SNe) \cite{BZ}, via shocks in the
interaction of their roughly spherical non-relativistic ejecta with the interstellar
medium (ISM) \cite{Shlovsky}. The NASA website lists almost 70000 refereed papers
on `cosmic ray(s)', most of them theoretical and
`standard'. As implied by a fraction of these papers, the standard theory has
problems: it does not accelerate CRs up to the knee (e.g.~\cite{Cesarsky}), 
the fraction of SN remnants perhaps compatible with the acceleration
of nuclear CRs is insufficient
to explain
the CR luminosity of the Galaxy (e.g.~\cite{HESS}), CRs produced by SN remnants
(most of which are within the `solar circle') would diffuse outwards,
 generating a directional asymmetry which is not observed
(e.g.~\cite{asymmetry}). CRs above the knee are mysterious.

Our theory \cite{DD2006} is {\it non-standard}  in that CRs
are accelerated {\it at all energies} by the {\it relativistic jets} ejected by  SNe. 
It is part of a `unified theory of high-energy
astrophysics' \cite{Florence}, 
based  on the `cannonball' model \cite{GRB1} of the jets of 
accreting black holes and neutron stars, which 
also explains simply
the properties of gamma-ray bursts (GRBs) \cite{DD}, 
X-ray flashes  (XRFs) \cite{DDDXRF}, their respective afterglows (AGs)
\cite{AGoptical,AGradio}, the gamma `background' radiation \cite{DDGBR}, the CR luminosity of our Galaxy \cite{DD2006,DDLum}, and the properties of
galaxy clusters harbouring `cooling flows'  \cite{CDD}. Many more and less
self-referring citations are given in \cite{DD2006,DD}.

\section{The Cannonball Model}

\begin{figure}
\vskip -1.cm
\hskip 2truecm
\begin{center}
{\epsfig{file=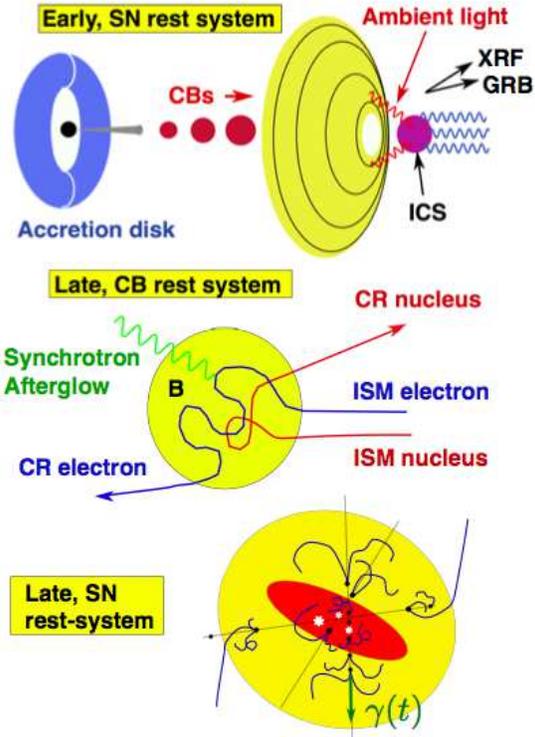, width=7.2cm}}
\end{center}
\vspace{-1.cm}
\caption{An `artist's view' (not to scale) of the CB model
of long-duration GRBs, XRFs and CRs. A core-collapse SN results in
a compact object and a fast-rotating torus of non-ejected
material. Matter (not shown) episodically accreting
into the central object produces
two narrowly collimated beams of CBs; only some of
the `northern' CBs are depicted. As the CBs move through
the `ambient light' surrounding the star, they forward Compton up-scatter
its photons to GRB or XRF energies, depending on how close
the line of sight is to the CBs' direction. 
Each CB produces a GRB `pulse'.
Later, a CB scatters ISM particles,
which are isotropized by its inner magnetic field. In the SN rest
system the particles are boosted by the CB's motion: they have become
CRs. The synchrotron radiation of the electrons is the late AG
of the GRB or XRF. The CBs generate CRs all along their 
trajectories, in the galaxy, its halo, and beyond, as the CBs'
collisions with the ISM
slow them down. CRs are also forward-produced, diffusing thereafter in the
local magnetic fields.}
\label{figCB}
\end{figure}

The `cannon' of the CB model is analogous to the ones
responsible for the ejecta of quasars and microquasars.
{\it Long-duration} GRBs, for instance, are produced in
{\it ordinary core-collapse} SNe by jets of CBs, made of {\it
ordinary-matter plasma}, and travelling with high LFs,
$\gamma\sim{\cal{O}}(10^3)$. As a consequence of the initial star's
rotation, an accretion disk  is produced around
the newly born compact object, either by stellar material originally
close to the surface of the imploding core, or by more distant stellar matter
falling back after the shock's passage \cite{GRB1,ADR}. A CB is emitted, as
in microquasars \cite{Felix}, when part of the accretion disk
falls abruptly onto the compact object. 
A summary of the CB 
model is given in Fig.~\ref{figCB}.

{\it Do SNe emit cannonballs?} Until 2003 \cite{030329}, there was only one
case with data good enough to tell: SN1987A, the core-collapse
SN in the LMC, whose neutrino emission was seen. Speckle interferometry
data taken 30 and 38 days after the explosion \cite{NP} 
did show two back-to-back relativistic CBs. The approaching one was `superluminal'.


{\it Are GRBs made by SNe?} For long-duration GRBs, the answer is 
affirmative \cite{AGoptical}. The first evidence for a  GRB--SN association 
concerned SN1998bw, 
at  redshift $z\!=\!0.0085$, observed  within the angular error 
towards GRB 980425.

GRBs have `afterglows': they are observable at 
radio to X-ray frequencies, for months after their $\gamma$-rays
are seen. The optical luminosity of a 1998bw-like SN 
peaks at $\sim\!15\,(1\!+\!z)$ days. The SN light 
competes at that time with the AG of its 
GRB: it is not always detectable.
 It makes sense {\it to test} whether long-duration GRBs 
are associated with a `standard torch' SN,  akin to SN1998bw, 
`transported' to their respective redshifts. The test works optimally:
{\it for all $\sim\!15$ cases in which such a SN could be seen, 
it was seen (with varying 
significance)} and {\it for all $\sim\!15$ cases in which the SN could not be seen,
it was not seen} \cite{AGoptical}. One could hardly do better. In practice SNe could 
not be observed at $z\!>\! 1.1$.

Naturally, truly `standard torches' do not exist, but SN1998bw
made such a good job of it that  we could {\it predict}~\cite{DDDSN,DDD329a}
the SN contribution to the AG in all  six recent cases of
early detection of the AGs of near-by GRBs.
 Besides 980425--1998bw,  
the most convincing  association was provided by the spectroscopic
discovery of a SN in the AG of GRBs 030329 
for which even the night when the SN 
would be discovered was foretold \cite{DDD329a}. 

In a CB-model analysis of GRBs and their 
AGs \cite{DD,AGoptical,AGradio} we find that,
within the pervasive cosmological 
factor of a few, the long-GRB--SN association would be $\sim$1:1.
Yet, 
current data are insufficient to determine whether long-duration
GRBs are associated with all core-collapse SNe ($\sim\!70$\% of all SNe,
including Type II) or only with Type Ib/c SNe ($\sim\!15$\% of core-collapse
SNe).

{\it CB-model priors.}
The study of GRB AGs \cite{AGoptical,AGradio} 
allowed us to extract, case by case, the initial
LF, $\gamma_0$, and (less precisely) the initial mass and
baryon number, $M_0\!\sim\! N_{_{\rm B}}\, m_p\, c^2$,
of CBs, as well as `environmental' 
quantities, such as the (highly varying) ISM density and the angle between 
the jet of CBs and the observer. Typical values are:
\begin{equation}
\gamma_0\equiv E/(M_0\,c^2)\approx(10^3);
\;\;\;\;
N_{_{\rm B}} \sim {\cal{O}}(10^{50}).
\label{typical}
\end{equation}
$M_0$
is half the mass of Mercury, i.e.~tiny,
compared to the mass of the parent  star. 
The $\gamma_0$ values are roughly log-normally distributed 
around $\gamma_0\!\sim\!1100$ with a width at half-maximum
extending from $\gamma_d\!\sim\!630$ to $\gamma_u\!\sim\!1400$ \cite{AGradio}.
With these inputs, we could predict all properties of the individual
$\gamma$ pulses of a GRB, each corresponding to a single CB
whose electrons Compton-up-scatter the ambient light \cite{DD}. 
The same inputs are used to predict the properties of CRs.

While a CB exits from its parent SN and emits a GRB pulse, it is
assumed~\cite{GRB1} to be expanding, in its rest system, at a speed close
to that of sound in a relativistic plasma ($v_s\!=\!c/\sqrt{3}$). CBs
continuously intercept electrons and nuclei of the ISM, ionized
by the GRB's $\gamma$-rays. Rapidly, such an expanding CB becomes 
`collisionless': its
radius becomes smaller than the interaction length between the constituents
of the CB and the ISM. But a CB
still interacts with the charged ISM particles, for it contains a
strong magnetic field.

Assume that the
ISM particles entering a CB's magnetic mesh are trapped,
slowly re-exiting by diffusion. Then, the CB's mass increases as:
\begin{equation}
M_{\rm CB}\approx 
M_0\,{\gamma_0/(\beta\,\gamma)},\;\;
\beta\equiv {\sqrt{\gamma^2-1}/ \gamma},
\label{NRmass}
\end{equation}
and, for an approximately hydrogenic ISM of local density $dn_{\rm in}$,
the CB's LF decreases as:
\begin{equation}
{d\,\gamma/(\beta^3\,\gamma^3)} \approx - \,{(m_p/M_0\,\gamma_0})\;
dn_{\rm in}(\gamma).
\label{gammadown}
\end{equation}

 We approximate a CB, in its rest system, by a sphere of radius $R(\gamma)$.
The ISM particles that are
intercepted ---isotropized in the CB's inner magnetic mesh,
and re-emitted--- exert an inwards force on it that counteracts its expansion. 
The computed behaviour of $R(\gamma)$ 
is well described by:
\begin{equation}
R_{_{\rm CB}}(\gamma)\approx  R_0\,
\left[{\gamma_0/(\beta\,\gamma)}\right]^{2/3},
\;\;\;\;
R_0\sim10^{14}\;{\rm cm}.
\label{best}
\end{equation}
The distances 
before a CB stops and blows up  range from
a fraction of a kpc to many kpc, depending on the density profile
they encounter.

The interactions of a CB and the ISM 
constitute the merger of two  plasmas at a large relative 
LF. This merger is very efficient in creating turbulent
currents and magnetic fields (MFs) within the CB.
We assume that the MFs,
as the CB reaches a quasi-stable radius, are in `equipartition': 
the MF energy density equals that
of the ISM particles the CB
has temporarily phagocytized.
This results in a MF~\cite{AGoptical}:
\begin{equation}
B_{_{\rm CB}}=3\;{\rm G}\;{(\gamma/ 10^3})\;
\left[{n_{\rm in}/ (10^{-3}\;{\rm cm}^{-3}})\right]^{1/2}\; ,
\label{B}
\end{equation}
where $n_{\rm in}$ is normalized to a typical value in the
`superbubbles' in which most SNe and GRBs are born.  This
result for $B(\gamma)$ is supported by the analysis of the
spectral evolution of GRB AGs.

Charged particles interacting with  turbulently moving MFs tend
to gain energy: a `Fermi' acceleration process. 
A `first-principle' numerical analysis~\cite{Fred} of
the merging of two plasmas
demonstrates the generation of such chaotic MFs,
and the acceleration of a small fraction of the injected
particles to the approximate spectrum:
\begin{eqnarray}
{dN/d\gamma_{_A}}&\propto&\gamma_{_A}^{-2.2}\,
\Theta(\gamma_{_A}-\gamma)\,\Theta[\gamma_{\rm max}(\gamma)-\gamma_{_A}],
\nonumber\\
\gamma_{\rm max}(\gamma) 
& \simeq & 10^5\;\gamma_0^{2/3}\;(Z/A)\; \gamma^{1/3},
\label{gammaA}
\end{eqnarray}
with $Z$ and $A$ the nuclear charge and mass.
The first  $\Theta$ function reflects the
fact that it is much more likely for the light particles to gain than to lose
energy in their elastic collisions with the heavy `particles' (the CB's
turbulent MF domains). The second $\Theta$  is 
the Larmor cutoff implied by the finite radius and
MF of a CB. But for the small dependence on the nuclear
identity (the factor $Z/A$), the spectrum of Eq.~(\ref{gammaA}) is
universal.

The average number 
of  pulses in a GRB's 
$\gamma$-rays  is $\sim 6$. Thus, the total energy of the two jets of CBs 
emitted by a core-collapse SN is:
\begin{equation}
E[{\rm jets}]\simeq 12\,\gamma_0\,N_{_{\rm B}}\,m_p\,c^2
\simeq 2\!\times\! 10^{51}\;\rm erg.
\label{Ejets}
\end{equation}
Practically all of this energy will, in our model, be transferred to CRs
and the MFs they produce.

Let $R_{\rm SN}\!\sim\!2$ per century be the SN rate in our Galaxy.
In a steady state, if the low-energy rays dominating the CR luminosity
are chiefly Galactic in origin, their accelerators must
compensate for the escape of CRs from the Galaxy.
The Milky Way's luminosity in CRs is then: 
\begin{equation}
L_{\rm CR}\!\approx\! 
R_{\rm SN}\,E{\rm[jets]}\!\approx\! 1.3\times 10^{42}\,\rm erg\, s^{-1}.
\label{SNEsupply}
\end{equation}

\section{Relativistic magnetic rackets}

The essence of our theory of CRs is kinematical and trivial.
In an {\it elastic} interaction of a CB at rest 
with ISM electrons or ions of LF $\gamma$,  
the light recoiling particles (of mass $m$) retain their incoming energy.
Viewed in the ISM rest system, they 
have a spectrum extending, for large $\gamma$,
up to $E\simeq 2\,\gamma^2\,m\,c^2$. A moving CB
is a gorgeous {\it Lorentz-boost accelerator:}
 the particles it elastically scatters with $\sim\!100$\% efficiency
reach up to, 
for $\gamma=\gamma_0\!=\!(1\,{\rm to}\,1.5)\!\times\!10^3$,
an $A$-dependent {\bf knee}  energy
$E_{{\rm knee}}(A)\!\sim\! (2\,{\rm to}\,4)\!\times\! 10^{15}\,A\,{\rm eV.}$

A particle of LF $\gamma$ entering a CB at rest may be
accelerated by elastic interactions with the CB's turbulent
plasma. Viewed in the rest system of the bulk of the CB, the interaction
is {\it inelastic} and the particle may re-exit with a LF up to
$\gamma_{\rm max}\!\sim\! 10^7\,\gamma$; see Eq.~(\ref{best}). 
Boosted by the CB's motion the spectrum of the scattered particles extends 
to $\gamma_{_{\rm CR}}\!\sim\! 2\!\times\! 10^7\,\gamma^2$, in the 
ultra-high-energy  (UHE)
 domain, for $\gamma\!\sim\gamma_0\!\sim 10^3$. 
This powerful {\it Fermi--Lorentz accelerator} completes our theory of CRs.

The calculation of the CR spectra
takes 2.6 pages \cite{DD2006}, which I do not have here.
For $\gamma\!>\! 2$,  to a good approximation, the elastic contribution 
to the CR flux of a nucleus of ISM abundance $n_{_A}$ is:
\begin{eqnarray}
{dF_{\rm elast}\over d\gamma_{_{\rm CR}}}&\propto& n_{_A}
\left({A\over Z}\right)^{\beta_{\rm conf}}
\int_1^{\gamma_0}{d\gamma\over \gamma^{7/3}}
\,{G[\gamma,\gamma_{_{\rm CR}}]}\; ,\nonumber\\
G[\gamma,\gamma_{_{\rm CR}}]&\equiv&
\int_{\rm max[\gamma,\gamma_{_{\rm CR}}/(2\,\gamma)]}
^{\rm min[\gamma_0,2\,\gamma\,\gamma_{_{\rm CR}}]}
{d\gamma_{\rm co}\over \gamma_{\rm co}^{4}}\; ,
\label{NRFlux}
\end{eqnarray}
where $\beta_{\rm conf}$ is the same `confinement' index as in
Eq.~(\ref{tururu});
$dF_{\rm elast}/d\gamma_{_{\rm CR}}$
depends on the priors $n_{_A}$, ${\beta_{\rm conf}}$, 
and $\gamma_0$, but not on any
parameter specific to the mechanism of CR acceleration.
The  inelastic contribution is equally simple \cite{DD2006}.

\begin{figure}
\begin{center}
\epsfig{file=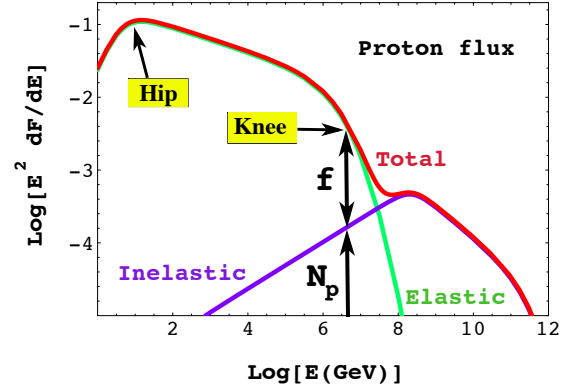, width=7.5cm,angle=-0}
\end{center}
\vspace{-1.2cm}
\caption{Contributions to the $E^2$-weighed proton source spectrum.
 Notice the predicted {\bf `hip'}.}
\vspace {-.8cm}
\label{DDelinel}
\end{figure}

The  source spectrum of a CR nucleus is the sum of an elastic and
an inelastic flux, illustrated, for protons, 
in Fig.~\ref{DDelinel}. The elastic flux is larger than the inelastic one
by a factor $f\!\simeq\! 10$ at the nominal position of the proton's knee. 
This ratio $f$ is the {\it only} input  for which we have
no `prior' information, and the only parameter to choose in an
unpredetermined range. 
The other parameter in Fig.~\ref{DDelinel}, ${N_p}$, is the norm
of the proton inelastic flux at the proton's knee. 
Albeit within large errors, $N_p$ (or $N_p\,f$) is determined from the 
luminosity of Eq.~(\ref{SNEsupply}), and the Galactic (or universal) rate of SNe.
In the domains wherein they behave as power laws, the 
elastic (inelastic) source spectra have indices $\beta_{\rm elast}\!=\!13/6$
($\simeq\!\beta_{\rm elast}+0.3$).

\begin{figure}
\vspace {-0.8cm}
\begin{center}
\epsfig{file=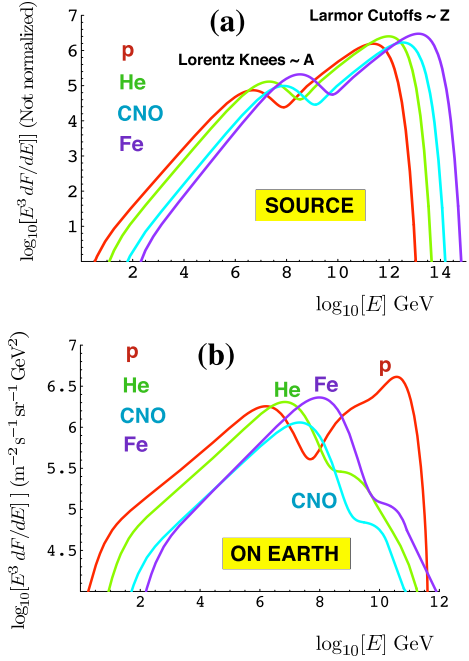, width=6.cm,angle=-0}
\end{center}
\vspace{-1.cm}
\caption{Predicted spectra for the abundant elements
and groups. The vertical scales are $E^3\,dF/dE$. (a): The source
spectra, with a common arbitrary normalization. (b): The CR spectra 
at the location of the Earth. Notice that both the horizontal and
vertical scales are different.}
\vspace{-.8cm}
\label{Groups}
\end{figure}

\nopagebreak
\section{Tribulations of a Cosmic Ray}
\label{tribulations}

The source and local fluxes of the main CR elements 
are shown in Fig.~\ref{Groups}. These fluxes differ because, on
its way from its source to the Earth, a CR 
is influenced by the ambient magnetic fields, radiation and matter.
An extragalactic CR is also affected by cosmological redshift. 
Three types of CR `tribulations' must be considered:

{\it Interactions with magnetic fields.} 
Fluxes of CRs of Galactic origin, below the free-escape 
({\bf ankle}) energy, $E_{\rm ankle}(Z)\!\sim\! Z\times (3\!\times\! 10^{18} \,\rm eV)$,
are enhanced proportionally to their
confinement time:
\begin{equation} 
\tau_{\rm conf}\propto \left[{Z\,{\rm GeV}/(c\, p)}\right]^{\beta_{\rm conf}},
\;
{\beta_{\rm conf}}\sim 0.6\pm 0.1. 
\label{tururu} 
\end{equation}
At higher energies CRs escape or enter the Galaxy practically
unhindered \cite{Cocconi}. Lower-energy
extragalactic CRs entering the Galaxy must overcome the effect of its 
exuding magnetic wind~\cite{DD2006}.

{\it Interactions with radiation,} significant for
CRs of extragalactic origin. The best studied one is the `GZK' 
$\pi$-photoproduction on the CMB.
Pair ($e^+ e^-$) production is analogous.
Photo-dissociation on the 
infrared CBR is also relevant. 

{\it Interactions with the ISM} are well understood. CR spallation gives rise
to `secondary' CRs.

\section{Results}

\begin{figure}
\vspace {-2.cm}
\begin{center}
\epsfig{file=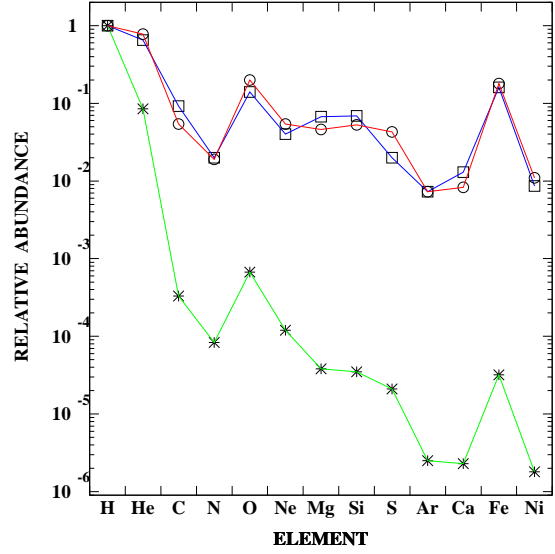, width=8.2cm,angle=0}
\end{center}
\vspace{-1.2cm}
\caption{The relative abundances of primary CR nuclei, from H to Ni
around 1 TeV.
The stars  are solar-ISM abundances. 
The circles  are the predictions, with input superbubble
abundances, a bit more `metallic' than the solar ones. The squares  are the 
CR observations.}
\label{f1}
\end{figure}

{\it Relative abundances.}
It is customary to discuss the composition of CRs at a fixed
energy $E_{_A}=1$ TeV.
This energy is relativistic, below the 
corresponding knees for all $A$, and in the domain wherein the 
fluxes are dominantly elastic and very well approximated by a power law 
of index $\beta_{\rm th}\!=\!\beta_{\rm elast}\!+\!\beta_{\rm conf}\!\simeq\!2.77$.
Expressed in terms of energy ($E_{_A}\!\propto\! A\,\gamma$), and modified 
by the confinement factor of Eq.~(\ref{tururu}), Eq.~(\ref{NRFlux}) becomes:
\begin{equation}
{dF_{\rm obs}/ dE_{_A}}\propto \bar{n}_{_A}\,A^{\beta_{\rm th}-1}
\,E_{_A}^{-\beta_{\rm th}},
\label{compo}
\end{equation}
with $\bar{n}_{_A}$ an average ISM abundance. 
At  fixed energy the prediction for the CR abundances relative to hydrogen is:
$X_{_{\rm CR}}(A)\!=\!(\bar{n}_{_A}/ \bar{n}_p)\,A^{1.77}$.
The results, for input $\bar{n}_{_A}$'s in the `superbubbles' wherein most
SNe occur, are shown in Fig.~\ref{f1}. 
Eq.~(\ref{compo})
snugly reproduces the large enhancements in the heavy CR abundances
relative to hydrogen, with respect to solar or superbubble abundances
(e.g.~$A^{1.77}\!=\! 1242$ for Fe).

\begin{figure}
\vspace {-1.cm}
\begin{center}
\epsfig{file=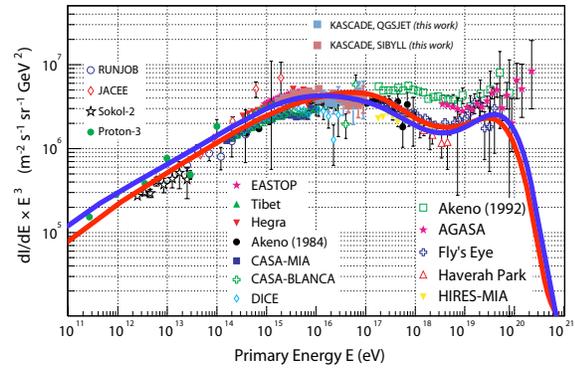, width=7.5cm,angle=-0}
\end{center}
\vspace{-1.2cm}
\caption{The all-particle CR spectrum. The vertical scale is 
$E^3\,dF/dE$. Some of the
high-energy data  disagree with others, and with our theory,
which opts for the results of fluorescence detectors.
The two curves correspond to two very different choices
for the `penetrability' of the Galaxy to extragalactic
CRs of energy below the ankle
and have slightly different values of two priors (both within errors):
the width of the $\gamma_0$ distribution and $\beta_{\rm conf}$, defined in
Eq.~(\ref{tururu}).}
\label{AllPart}
\end{figure}

{\it The all-particle spectrum} is shown in  
Fig.~\ref{AllPart}. 
The curves in this figure (and in later ones)
correspond to  two very different choices of the Galaxy's `penetrability'
 to extragalactic CRs \cite{DD2006}.

\begin{figure}
\vspace {-.8cm}
\centering
\epsfig{file=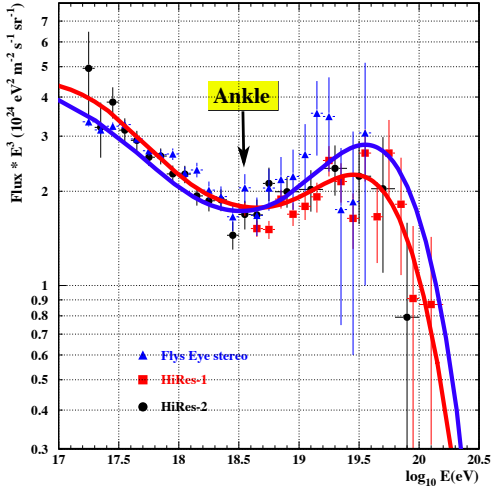, height=7cm,  width=7cm,angle=-0}
\vspace{-1.cm}
\caption{The $E^3$-weighed  UHECR spectrum \cite{HIRES1}. 
The colour coding is as in Fig.~\ref{AllPart}.}
\label{UHECR}
\end{figure}

The  UHECR all-particle spectrum is
shown in Fig.~\ref{UHECR}. At $E=E_{\rm ankle}$ the extragalactic
contribution is $\sim\!1/2$ of the data.
 The flux shape and norm, at this energy or above it, are
approximate but `absolute': they are the `look-back-time' integral of the CR flux due
to SNe in other galaxies.
The shape of the high-energy end-point 
and the height of the hump reflect not only the GZK cutoff, 
but also the `Larmor' cutoff for protons. 

\begin{figure}[]
\centering
\vskip .5cm
\vbox{\epsfig{file=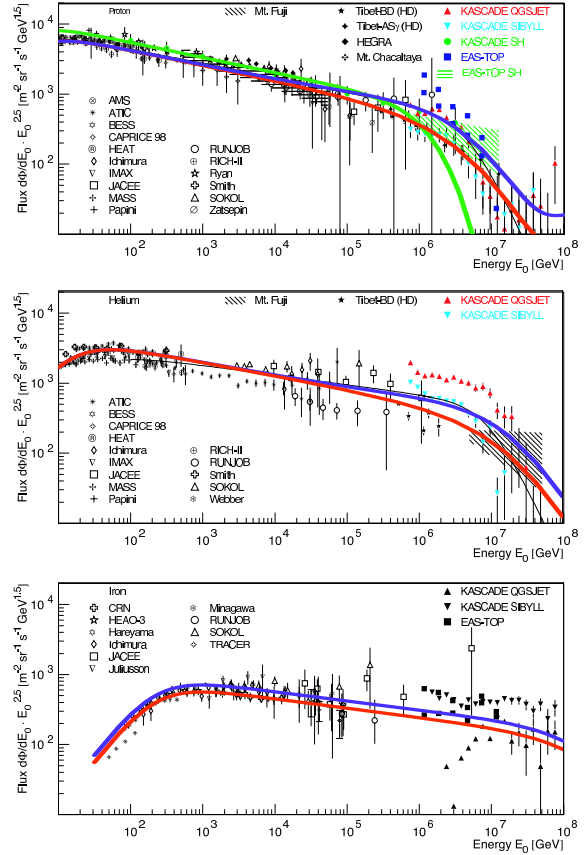,width=7.5cm}}
\caption{Measurements of individual-element CR spectra in
the `knee' region \cite{Hoer}. The vertical scales are $E^{2.5}\,dF/dE$.
Top: protons; middle: $\alpha$ particles; bottom: iron nuclei.
The data were kindly provided to us by K.H. Kampert. The colour
coding is as in Figs.~\ref{AllPart} and \ref{UHECR}. The green line in the
proton entry corresponds to a narrow prior $\gamma_0$-distribution.
The $\alpha$ data and, more so, the Fe data, show the predicted
low-energy `hips', see also Figs.~\ref{DDelinel} and \ref{VeryLowEnergy}.
Notice the dependence of the KASKADE results on the chosen
`shower MonteCarlo'. The data are insufficient to tell apart 
$E_{\rm knee}\!\propto\!A$ (our prediction) from $E_{\rm knee}\!\propto\!Z$
(sometimes quoted as the standard-theory result, but correct {\it iff} the limit is
a Larmor cutoff and CR nuclei were accelerated up to their knees).}
\label{KASKADE}
\end{figure}

{\it The knee region.}
There are recent KASKADE data attempting
to disentangle the spectra of individual elements or groups in the
knee region. 
Our predictions
for the spectra of  H, He and Fe are shown in Fig.~\ref{KASKADE}.
The green line in the proton entry has a narrow 
$\gamma_0$ distribution, the 
red and blue lines have wider distributions (all within the errors
in this `prior') and correspond to the equally coloured lines in
Figs.~\ref{AllPart} and \ref{UHECR}.
%

\begin{figure}
\vspace {-0.5cm}
\begin{center}
\epsfig{file=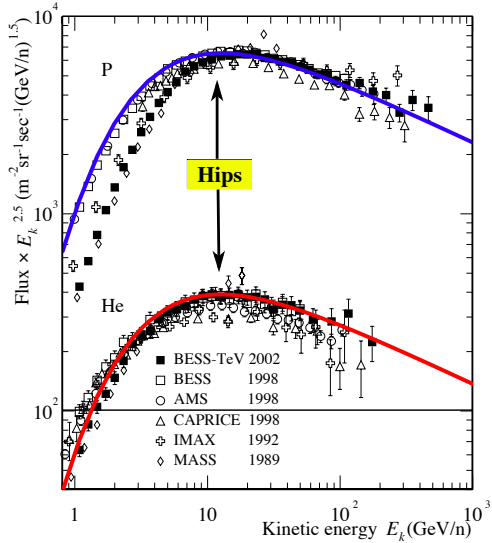, width=7cm,angle=-0}
\end{center}
\vspace{-1.5cm}
\caption{The very-low-energy fluxes of protons and $\alpha$ particles
at various times in a solar cycle. The 1998 data are close to solar-minimum
time.}
\label{VeryLowEnergy}
\end{figure}

{\it The low-energy spectra.}
In Fig.~\ref{VeryLowEnergy} we show
the weighted spectra $E_k^{2.5}\,dF/d\,E_k$
of protons and $\alpha$ particles, as functions of $E_k$, the 
kinetic energy per nucleon. The data were taken at
various times in the 11-year solar cycle. The most intense
fluxes correspond to data near a solar minimum.
The curves do not  model the
effects of the solar wind. They should agree best with the
solar-minimum data, as they do, particularly for protons.
The theoretical source spectra, dominated by the elastic contribution,
 are given by Eq.~(\ref{NRFlux}). These data are well below the elastic cutoff
at $\gamma_{_{\rm CR}}\simeq 2\,\gamma_0^2$, and the predictions are 
independent of the  $\gamma_0$ distribution. Thus, the shape 
of the theoretical spectra is, in this energy domain, parameter-free.
%

\begin{figure}
\vspace {0.5cm}
\hbox{\hskip 0.7cm\epsfig{file=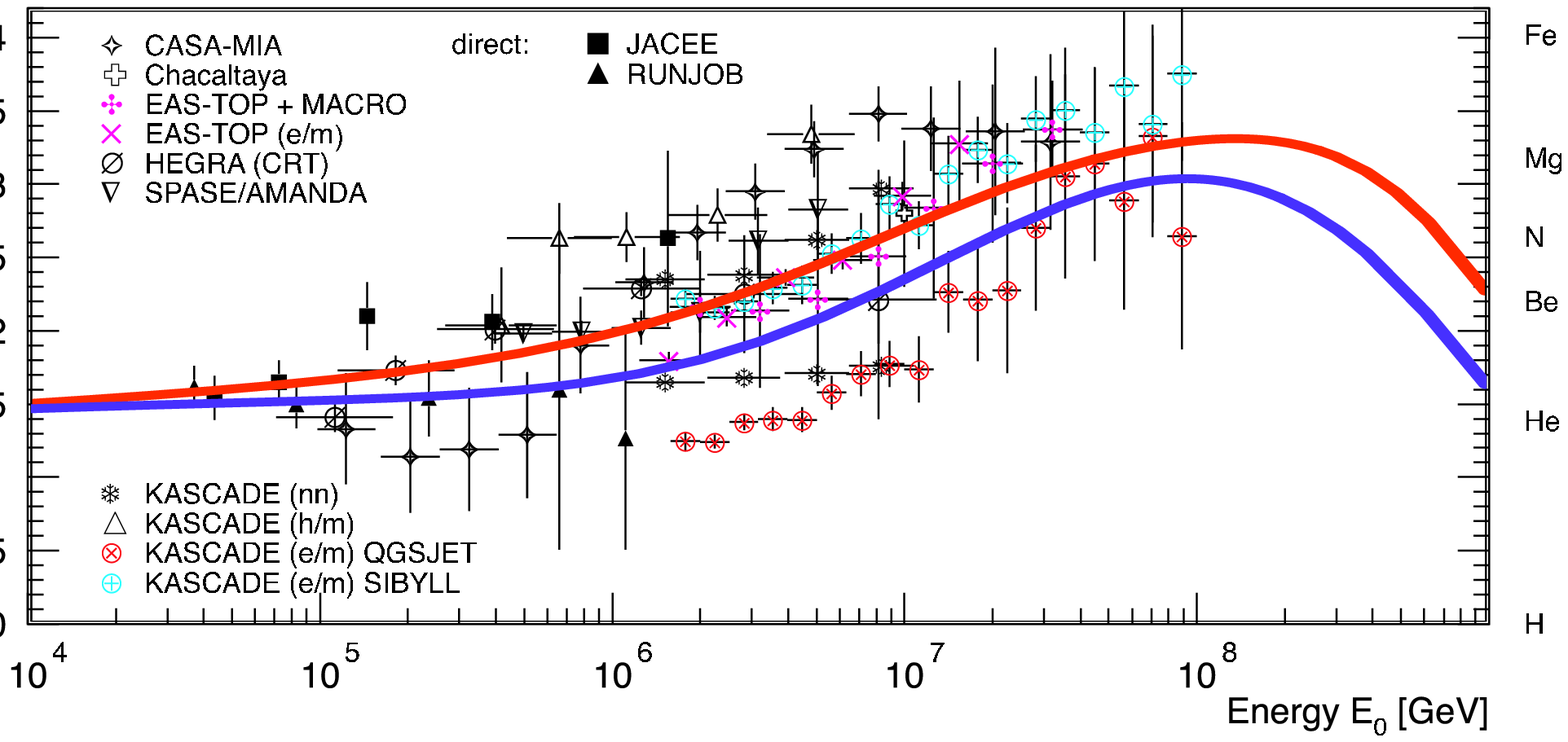, width=6.8cm,angle=-0}}
\vspace {-.5cm}
\caption{Mean logarithmic mass of  CRs. Data points were
         compiled  by Hoerandel \cite{Hoer} from
         experiments measuring electrons, muons, and hadrons at
         ground level. The colour-coded lines correspond to the same choices 
of priors as in Figs.~\ref{AllPart}--\ref{KASKADE}.
The compilation of data was kindly provided to us by K.H. Kampert.}
\label{lnAknee}
\end{figure}

\begin{figure}
\begin{center}
\epsfig{file=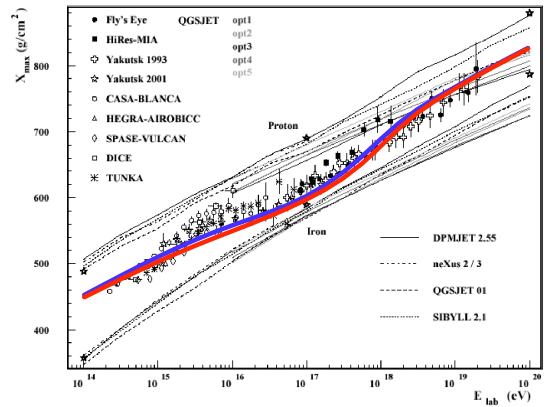, width=8.5cm,angle=-0}
\end{center}
\vspace{-7cm}
\caption{The depth of shower maximum as a function of energy. The data are from a
compilation in Ref.~\cite{ZKP}. The colour-coded lines
correspond to the same choices as in 
Figs.~\ref{AllPart}--\ref{KASKADE} and \ref{lnAknee}.
The theoretical lines are constructed with the simplified method of
Wigmans \cite{Wigmans}, arguably as good as any currently-used MonteCarlo.}
\label{Xmax}
\end{figure}

{\it Rough measures of CR composition.}
The evolution of the CR composition with energy is
often presented in terms of the mean logarithmic
atomic weight $\langle \ln A \rangle$, or of the depth into the
atmosphere of the `maximum' of the CR-generated shower,
$X_{\rm max}$. The predicted   $\langle \ln A(E) \rangle$ is
compared with relatively low-energy
data in Fig.~\ref{lnAknee}. Results for $X_{\rm max}(E)$,
are shown in Fig.~\ref{Xmax}.

\begin{figure}
\vspace {0.5cm}
\begin{center}
\epsfig{file=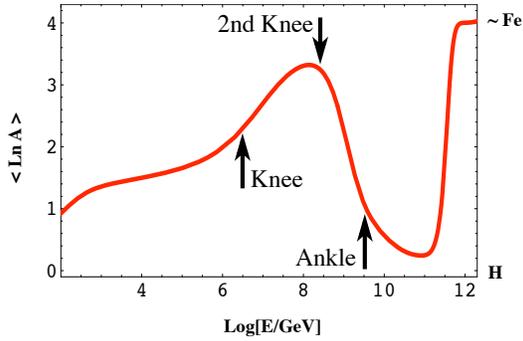, width=7cm,angle=-0}
\end{center}
\vspace{-1.2cm}
\caption{Predicted $\langle \ln A(E) \rangle$ at all energies.}
\label{lnA}
\end{figure}

The predicted $\langle \ln A(E) \rangle$ at all energies, 
shown in Fig.~\ref{lnA}, shows how, at very high energies,
the flux is once more Fe-dominated: lighter elements have
reached their GZK, acceleration and Galactic-escape cutoffs.
Naturally, this prediction is  sensitive to the 
details of Galactic escape and extragalactic photodissociation.

\begin{figure}[]
\centering
 \epsfig{file=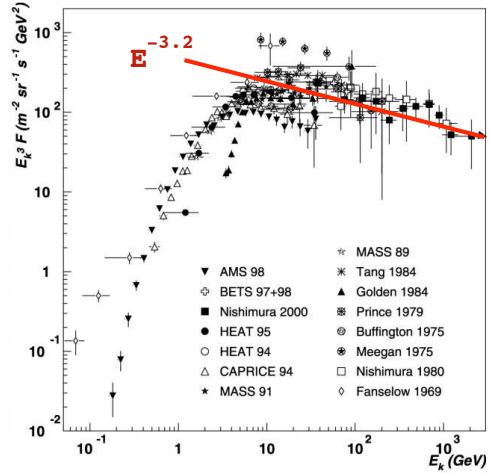,width=6.5cm}
\vspace{-1cm}
\caption{The CRE spectrum, compiled in  \cite{Sapinski}.
The line is the central result of a power fit to the higher-energy data; 
its slope is $3.2\pm0.1$ \cite{Sreek}. The different slope below 
$\sim\!6$ GeV results from the diffusion of electrons
in the Galactic MF  \cite{DD2006}.}
 \label{CREspectrum}
\end{figure}

{\it Cosmic-ray electrons.}
Electrons and nuclei are accelerated by the `magnetic-racket'
CBs in the same manner \cite{DD2006,DDGBR}. 
The functional form of their source spectra is
therefore the same,
$dF_s/d\gamma\propto\! \gamma^{\beta_{\rm elast}}$, 
in the range $10\!<\!\gamma\!<\!10^6$.
Electrons lose energy much more
efficiently than protons in their interactions with radiation, MFs and the ISM.
The rates  $-dE/dt\!\propto\! E^{\alpha}$ of their various 
mechanisms of energy loss have different $\alpha$'s. 
For Coulomb losses $\alpha=0$; for bremsstrahlung $\alpha=1$; for 
inverse Compton scattering (ICS)
and synchrotron losses at the relevant energies, $\alpha=2$. 
At sufficiently high energy, the radiative energy loss 
dominates the others.
In this domain, the steady-state
solution of the equation describing the radiation-modified electron spectrum,
for a source $dF_s/dE\propto E^{-\beta_{\rm elast}}$, is simply \cite{Anton,DDGBR}:
\begin{equation}
{dF_e/ dE}\propto E^{-\beta_e};\;\;\; \beta_e=\beta_{\rm elast}+1\approx 3.17.
\label{espectr}
\end{equation}
This result agrees with the observed slope of
the CRE spectrum; see Fig.~\ref{CREspectrum}.
The best-fitted value above $E\!\sim\!6$ GeV is
$\beta_{\rm obs}=3.2\pm 0.10$, and the
fit is excellent if the experiments are recalibrated to the same 
flux at high energy \cite{DDGBR}.

\begin{figure}[]
\vspace{-.5cm}
\centering
 \epsfig{file=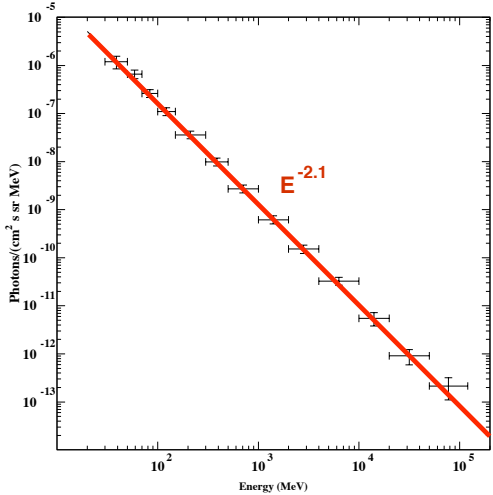,width=7.2cm}
\vspace{-1cm}
\caption{The GBR spectrum, measured by EGRET \cite{Sreek}.
The line is the central result of a power-law fit of
slope  $2.10\pm0.03$.}
 \label{GBRspectrum}
\end{figure}

{\it The GBR.} The existence
of a diffuse gamma background radiation,
suggested by data from the SAS 2 satellite,
was confirmed by the
EGRET instrument on the Compton Gamma Ray Observatory \cite{Sreek}.  
We call `the GBR' the diffuse emission observed by EGRET
by masking the galactic plane at latitudes
$\rm{|b|\le 10^o}$, as well as the galactic centre
at $\rm{|b|\le 30^o}$ for longitudes $\rm{|l|\le 40^o}$,
and by extrapolating to zero column density, to eliminate the $\pi^0$
and bremsstrahlung contributions to the observations and
to tame the model-dependence of the results.
Outside this `mask', the GBR flux integrated over all directions, 
shown in
Fig.~\ref{GBRspectrum}, is well described by a power law 
$dF_\gamma/dE\!\propto\! E^{-2.10\pm 0.03}$ \cite{Sreek}.

The EGRET GBR data show a significant deviation from isotropy, correlated with the
structure of the Galaxy and our position relative to its centre \cite{Anton}.
Contrariwise, the GBR's spectral index is uncannily directionally uniform.
These facts suggest a GBR that is partially local, as opposed
to dominantly cosmological, and a common origin for the
Galactic and extragalactic contributions.

In \cite{Anton,DDGBR} we have analyzed the directional and spectral 
properties of the EGRET data and concluded that the
GBR is produced by ICS of CREs on starlight and the CBR. It has
comparable contributions from CREs in a Galactic halo
of dimensions akin to the hadronic-CR confinement volume
(a directional and local source) and from other galaxies 
(an isotropic cosmological component).
Thus, the GBR is a CR `secondary'. 
Its spectral index is the same for the local and cosmological contributions.

%
If produced by ICS by electrons with the spectrum of
Eq.~(\ref{espectr}), the GBR has a spectrum:
\begin{equation}
{dF^i_\gamma/ dE}\propto E^{-\beta_\gamma},\;\;\;
\beta_\gamma={(\beta_e-1)/2 }\simeq 2.08.
\label{ICSphotpred}
\end{equation}
 The predicted photon spectral index \cite{Anton,DDGBR}
coincides with the measured one, ${2.10\pm 0.03}$ \cite{Sreek}.

{\it Other predictions.}
Our CR theory explains other observations that I have no space to
discuss here: the slight differences between the slopes of the
spectra of the hadronic CRs, the deduced confinement time and volume
of CR electrons and nuclei in the Galaxy \cite{DD2006}, and the 
normalization and 
directional dependence of the GBR flux \cite{DDGBR}.

\section{Discussion and conclusions}

Our theory is incomplete in several respects. The 
ejection of CBs in episodes of accretion onto a compact object is
supported by observations, but not fundamentally understood
(this is also the case for the ejection of SN shells). Because
CBs deposit CRs along their kpc-long trajectories, CR diffusion
does not play a crucial role, but we have not studied it in detail
(in the standard theory diffusion results in directional asymmetries
that are not observed, and calls for ad-hoc remedies, such as
`CR reacceleration'). The temporary confinement of CRs ---in 
the Galaxy or within a CB--- is not fully understood. Neither is the
`penetrability' of the Galaxy to extragalactic CRs of energy below
the ankle. We have studied the dynamics of the 
expansion of CBs, but not modelled it in minute detail. We contend
that a good fraction of the original energy of CRs ends up in the
production of  `equipartition' MFs~\cite{DDMF}, but we cannot predict this, nor
determine the effect that it may have on the CR spectral shape.

In spite of the above limitations, we have 
demonstrated how our simple {\it and single}
accelerators ---cannonballs--- are effective at all observed 
energies. The mechanisms of CR acceleration, particularly
the `elastic' one, could hardly be simpler: a fast massive object
slows down by kicking out of its way the light particles
it encounters. 

Most of our results are `robust' in that ---within very large brackets---
they do not depend on the specific choices of parameters and priors:\\
$\bullet$ An all-particle piecewise power-law spectrum with four 
features: two steepenings at the knee and the second knee, a
softening at the ankle, and an end-point at the roughly-coincident
GZK and proton-acceleration cutoffs.\\ 
$\bullet$ An UHECR flux above the ankle, which is predicted ---to 
within a factor of a few--- and otherwise parameter-free.\\
$\bullet$ A composition dependence at 1 TeV with the observed trend,
so different from that of the ISM.\\ 
$\bullet$ A very low-energy flux whose spectral shape is independent of any 
CB-model `prior' parameters.\\
$\bullet$ Individual-element knees that scale like $A$ and occur at the 
predicted energies.\\
$\bullet$ A non-trivial shape of the individual knees: an abrupt decrease in flux,
followed by a spectrum 
 steeper than that below the knee.\\
$\bullet$ An ankle with the observed shape. The dominantly Galactic-Fe flux
below it and the dominantly extragalactic-proton flux above it are comparable
in magnitude at the estimated escape `ankle' energy \cite{Cocconi} of Galactic protons.\\
$\bullet$ A composition dependence that is almost energy-independent below
the knee becomes `heavier' from the knee to the second knee,
`lighter' again above it, and finally heavier at yet unmeasured ultra-high
energies.\\
$\bullet$ An `extended' distribution of CR sources along CB trajectories emerging 
from the central realms of the Galaxy, where most SN explosions occur, 
implying a CR flux at the Earth's location with a much 
smaller and less energy-dependent anisotropy than that of standard
 models of CRs.\\
$\bullet$ 
Predictions for the values of a related set of observables:
the CR luminosity, confinement time and volume of the Galaxy, the spectral
indices of CR electrons and of the diffuse GBR. \\

Our results describe the observed properties of hadronic non-solar
CRs very well from the lowest energies  to
$\sim 10^{10}$ GeV. Above that energy and up to the highest observed
energies, $\sim 10^{11}$ GeV, our theory opts for the data gathered with
fluorescence detectors. Overall, the energy range for which the theory is
successful covers ten decades
and the flux extends over  three times as many.



%

\section*{Acknowledgements}
I thank Shlomo Dado and Arnon Dar for a long and fruitful collaboration, and
Nick Antoniou, Giuseppe Cocconi,
Andy Cohen, Sergio Colafrancesco, Shelly Glashow and
Rainer Plaga for collaboration and/or patient discussions.

\end{document}